\documentclass[fleqn,usenatbib]{mnras}

\usepackage{newtxtext,newtxmath}

\usepackage[T1]{fontenc}
\usepackage{ae,aecompl}
\usepackage{soul}
\usepackage{graphicx}	% Including figure files
\usepackage{amsmath}	% Advanced maths commands
\usepackage[usenames]{xcolor}
\usepackage{natbib}
\usepackage{hyperref}
\usepackage{xfrac}

\def\equationautorefname~#1\null{equation~(#1)}

\DeclareMathAlphabet{\mathpzc}{OT1}{pzc}{m}{it}\definecolor{purple}{RGB}{160,32,240}
\newcommand{\nick}[1]{\textcolor{black}{ #1}}
\newcommand{\Msun}{M_{\odot}}
\newcommand{\Mstar}{M_{\star}}
\newcommand{\Mearth}{M_{\oplus}}
\newcommand{\Mj}{M_{\rm J}}

\newcommand{\Sigg}{\Sigma_{\rm gas}}
\newcommand{\Sigd}{\Sigma_{\rm dust}}
\newcommand{\gcm}{\rm g/cm^2}
\title[Accreting ALMA Protoplanets]{Testing planet formation 
from the ultraviolet to the millimeter}

\author[Choksi \& Chiang]{Nick Choksi$^{1 \href{https://orcid.org/0000-0003-0690-1056}{\includegraphics[scale=0.4]{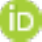}}}$\thanks{E-mail: nchoksi@berkeley.edu} and
Eugene Chiang$^{1,2\href{https://orcid.org/0000-0002-6246-2310}{\includegraphics[scale=0.4]{figures/orcid.pdf}}}$
\\
$^{1}$Astronomy Department, Theoretical Astrophysics Center, and Center for Integrative Planetary Science, University of California\\
\hspace{0.015in} Berkeley, Berkeley, CA 94720, USA\\
$^{2}$Department of Earth and Planetary Science, University of California, Berkeley, CA 94720, USA
}

\date{Released \today}

\pubyear{2021}

\begin{document}
\label{firstpage}
\pagerange{\pageref{firstpage}--\pageref{lastpage}}
\maketitle

\begin{abstract}
Gaps imaged in protoplanetary discs are suspected to be 
opened by planets. We 
compute the present-day mass accretion rates $\dot{M}_{\rm p}$ of seven hypothesized gap-embedded planets, plus the two confirmed planets in the PDS 70 disc. The accretion rates are based on disc gas surface densities $\Sigma_{\rm gas}$ 
from C$^{18}$O observations, and planet masses $M_{\rm p}$ from simulations fitted to observed gaps. Assuming accretion is Bondi-like, we find in eight out of nine cases that $\dot{M}_{\rm p}$
is consistent with the time-averaged value given by the current planet mass and system age, $M_{\rm p}/t_{\rm age}$. As system ages are comparable to circumstellar disc lifetimes, these gap-opening planets 
may be undergoing their last mass doublings, reaching final masses of $M_{\rm p} \sim 10-10^2 \, M_\oplus$ for the non-PDS 70 planets, and $M_{\rm p} \sim 1-10 \, M_{\rm J}$ for the PDS 70 planets. 
For another fifteen gaps without C$^{18}$O data, we predict $\Sigg$ by assuming their planets are accreting at their time-averaged $\dot{M}_{\rm p}$. Bondi accretion rates for PDS 70b and c are orders of magnitude higher than accretion rates implied by measured U-band and H$\alpha$ fluxes, suggesting most of the accretion shock luminosity emerges in as yet unobserved wavebands,
or that the planets are surrounded by dusty, highly extincting, quasi-spherical circumplanetary envelopes. Thermal emission from such envelopes or from circumplanetary discs, on Hill sphere scales, peaks at wavelengths in the mid-to-far-infrared and can reproduce observed mm-wave excesses.
\end{abstract}


%%%%%%%%%%%%%%%%%%%%%%%%%%%%%%%%%%%%%%%%%%%%%%%%%%
%%%%%%%%%%%%%%%%% BODY OF PAPER %%%%%%%%%%%%%%%%%%

\section{Introduction}
\label{sec:Intro}

The Atacama Large Millimeter Array 
(ALMA) has revealed gaseous protoplanetary discs to be ringed and gapped, with HL Tau (\citealt{alma_etal_2015}) and discs from the DSHARP  \citep{andrews_etal_2018}
and MAPS \citep{oberg_etal_2021}
surveys providing stunning examples (see also \citealt{cieza_etal_2019, andrews_etal_2020}). The annular structures observed at orbital distances of $r \sim 10-100$ au are commonly interpreted as being carved by protoplanets within the gaps, by analogy with shepherd satellites opening gaps in planetary rings (e.g. \citealt{goldreich_tremaine_1979}). A few dozen candidate planets have been so hypothesized with masses $M_{\rm p}$ of several Earth masses $\Mearth$ to a few Jupiter masses $\Mj$  (\citealt{zhang_etal_2018}; figure 1 of \citealt{lodato_etal_2019}; and Table \ref{tab:data} of this paper).

So far disc-embedded protoplanets have been directly imaged in one system: PDS 70b and c are confirmed in the near-infrared to orbit at $r \sim 20-40$ au within the central clearing of the system's transitional disc (\citealt{keppler_etal_2018}; \citealt{haffert_etal_2019}; \citealt{wang_etal_2020, wang_etal_2021}, and references therein).
These protoplanets, each estimated to be of order a Jupiter mass, have been imaged at multiple wavelengths, from the ultraviolet at 336 nm \citep{zhou_etal_2021} to H$\alpha$ \citep{haffert_etal_2019} to the sub-millimeter continuum \citep{isella_etal_2019, benisty_etal_2021}. Aside from PDS 70b and c, all other planets proposed to explain disc sub-structures remain undetected. There are unconfirmed near-infrared point sources in HD 163296 (\citealt{guidi_etal_2018}) and Elias 24 (\citealt{jorquera_etal_2021}), and kinematical evidence for
planetary perturbations to the rotation curve in HD 163296 (\citealt{teague_etal_2018, teague_etal_2021}).

Are the disc gaps and rings imaged by ALMA
and in the infrared (e.g. \citealt{asensio-torres_etal_2021}) really caused by planets? What can we hope to learn from DSHARP discs and similar systems about how planets form? At the top of the list of measurables should be the planetary mass accretion rate $\dot{M}_{\rm p}$ and how it depends on $M_{\rm p}$ and ambient disc properties like density and temperature. The planet masses inferred from modeling disc sub-structures are frequently dozens of Earth masses (e.g. \citealt{dong_fung_2017}; \citealt{zhang_etal_2018}; our Table \ref{tab:data}),\footnote{The gaps simulated by \citet{zhang_etal_2018} and similar works are opened  because of the repulsive Lindblad torques exerted by planets. These planet-disk simulations neglect accretion onto the planet (cf.~\citealt{rosenthal_etal_2020} and references therein).} large enough that an order-unity fraction of their mass should be in the form of accreted hydrogen. For 
such planets with self-gravitating gas envelopes, accretion may be hydrodynamic, i.e. 
regulated by how much mass the surrounding nebula can deliver to the planet through transonic/supersonic flows
(e.g. \citealt{dangelo_etal_2003}; \citealt{machida_etal_2010};  \citealt{tanigawa_tanaka_2016};  \citealt{ginzburg_chiang_2019a}). 
This phase is distinct from an earlier  thermodynamic phase of gas accretion limited by the rate at which the nascent envelope can cool and contract quasi-hydrostatically (e.g. \citealt{elee_chiang_2015}; \citealt{ginzburg_etal_2016}; \citealt{chachan_etal_2021}).
Are the inferences of planets still immersed in their parent discs consistent with our ideas of hydrodynamic accretion? To what final masses should we expect the hypothesized planets to grow?

This paper takes stock of the developing landscape of disc sub-structures and directly imaged protoplanets to address these and other questions. Our goal is to assess what we have learned (if anything) about planet formation from the multi-wavelength observations, and how theory fares against data. In these early days, the analysis is necessarily at the order-of-magnitude level. We begin in Section \ref{sec:data} by documenting our sources (mostly DSHARP) for disc and putative planet properties. These quantities are used in Section \ref{sec:results} to estimate the present-day $\dot{M}_{\rm p}$ according to theory and to evaluate whether the corresponding planet growth timescales $M_{\rm p}/\dot{M}_{\rm p}$ make sense in comparison to the system ages. The mass accretion rates are also used to evaluate accretion luminosities, and we discuss how accretion may be playing out in the circumplanetary environments of PDS 70b and c.
In Section \ref{sec:discussion}, we summarize and offer an outlook.

\begin{table*}
\centering
\begin{tabular}{|l|c|c|c|c|c|c|l|l|}
\hline
\\[-2mm]
\,\,\,\,(1) & (2) & \,\,(3) & (4) & (5) & (6) & (7) & \,\,(8) \\
 Name &  $M_\star$ [$M_{\odot}$]  &   $t_{\rm age}$ [Myr] & $r$ [au] & $M_{\rm p}$ [$\Mearth$] & $\Sigg$ [g cm$^{-2}$] & $\Sigma_{\rm dust}$ [g cm$^{-2}$]  & $h/r$ & 
 \\ \hline
 Sz 114 & 0.17 & 1 & 39 & $3-6$ &  N/A   & $0.08 - 0.6$ &  0.1 &  \\ 
 GW Lup & 0.46 & 2 & 74 & $2-10$ &  N/A   & $0.03 - 0.2^{\dagger}$ &  0.08  &  \\ 
 Elias 20 & 0.48 & $\leq$ 0.8 & 25 & $8-20$ &  N/A   & $0.1 - 0.7$ &  0.08 &  \\ 
 Elias 27 & 0.49 & 0.8 & 69 & $3-20$ &  N/A   & $0.05 - 0.3$ &  0.09  &  \\ 
 RU Lup & 0.63 & 0.5 & 29 & $10-20$ &  N/A   & $0.2 - 1$ &  0.07 &  \\ 
 SR 4 & 0.68 & 0.8 & 11 & $60-700$ &  N/A   & $0.03 - 0.2^{\dagger}$ &  0.05 &  \\ 
 Elias 24 & 0.78 & 0.2 & 57 & $30-300$ &  N/A   & $0.003 - 0.02$ &  0.09  &  \\ 
TW Hya-G1 & 0.80 & 8 & 21 & $8-80$ &  $0.04-3$   & $0.06 - 0.4$ &  0.08$^{\ast}$  & \\ 
TW Hya-G2 & 0.80 & 8 & 85 & $5-50$ &  $0.008-0.2$   & $0.002 - 0.02$ &  0.09$^{\ast}$  & \\ 
Sz 129 & 0.83 & 4 & 41 & $5-10$ &  N/A   & $0.2 - 1$ &  0.06 &  \\  
DoAr 25-G1 & 0.95 & 2 & 98 & $20-30$ &  N/A   & $0.05 - 0.3$ &  0.07 &  \\ 
DoAr 25-G2 & 0.95 & 2 & 125 & $5-10$ &  N/A   & $0.08 - 0.6$ &  0.07 &  \\ 
IM Lup & %0.89 
1.1 & 0.5 & 117 & $10-30$ &  $0.1-10$   & $0.03 - 0.2$ &  0.1$^{\ast}$ &  \\ 
AS 209-G1 & 1.2%0.83 
& 1 & 9 & $60-700$ &  N/A    &  $0.05-0.3^{\dagger}$ &  0.04  &  \\ 
AS 209-G2 & %0.83 
1.2 & 1 & 99 & $30-200$ &  $0.04-0.4$    &  $0.008 - 0.06$ &  0.06$^{\ast}$  &  \\
HD 142666 & 1.58 & 10 & 16 & $10-100$ &  N/A   & $0.2 - 1$ &  0.05  &  \\ 
HD 169142 & 1.65 & 10 & 51 & $30-300$ & $0.1-0.2$   & $\leq 0.01$ &  0.07$^{\ast}$ &  \\ 
HD 143006-G1 & 1.78 & 4 & 22 & $400-6000$ &  N/A   & $0.003 - 0.02^{\dagger}$ &  0.04  &  \\ 
HD 143006-G2 & 1.78 & 4 & 51 & $20-100$ &  N/A   & $0.02 - 0.2$ &  0.05  &  \\ 
HD 163296-G1 & 2.0 & 10 & 10 & $30-200$ &  N/A   & $0.2 - 1$ &  0.07$^{\ast}$  &  \\ 
HD 163296-G2 & 2.0 & 10 & 48 & $90-700$ &  $1-40$   & $0.003 - 0.02^{\dagger}$ &  0.08$^{\ast}$  &  \\ 
HD 163296-G3 & 2.0 & 10 & 86 & $10-300$ &  $0.1-20$   & $0.01 - 0.09$ &  0.08$^{\ast}$  &  \\ 
\hline 
PDS 70b & 0.88 & 5 & 22 & $300-3000$ & $0.0008 - 0.08$ & $\leq 0.02$ & 0.07$^{\ast}$ \\ 
PDS 70c & 0.88 & 5 & 34 & $300-3000$ & $0.0008 - 0.08$   & $\leq 0.02$ & 0.08$^{\ast}$ \\ \hline 
\end{tabular}
\caption{Properties of gapped discs and the planets hypothesized to reside within them (confirmed in the case of PDS 70). Column headings: (1) System name. For systems with multiple gaps, we append ``G\#'' to distinguish between different gaps. (2) Stellar mass \protect 
\citep{andrews_etal_2018,  ducourant_etal_2015,oberg_etal_2021,blondel_dije_2006,   keppler_etal_2019} (3) Stellar age \protect \citep{andrews_etal_2018,ducourant_etal_2015, pohl_etal_2017, muller_etal_2018} (4) Planet orbital radius 
\protect \citep{zhang_etal_2018, dong_fung_2017, keppler_etal_2018} 
(5) Planet mass. For the non-PDS 70
planets (first 22 entries), the quoted range in $M_{\rm p}$ corresponds to disc viscosities $\alpha = 10^{-5}-10^{-3}$, scaled from the gap-fitting, two-fluid, planet-disc simulations of \protect \citet{zhang_etal_2018} and \citet{dong_fung_2017} (see section \ref{subsubsec:masses_dsharp}). 
The ranges on $M_{\rm p}$ for PDS 70b and c roughly cover values inferred from infrared photometry and cooling models \protect \citep{muller_etal_2018, keppler_etal_2018, wang_etal_2020}. (6) Gas surface density at the planet's location, as inferred from C$^{18}$O emission \protect \citep{  nomura_etal_2021,zhang_etal_2021, favre_etal_2019, fedele_etal_2017, isella_etal_2016, facchini_etal_2021}. The quoted ranges on $\Sigg$ in AS 209, IM Lup, HD 163296, and TW Hya accommodate the possibility that the CO:H$_2$ abundance ratio in these discs is lower than in the ISM  \protect \citep{zhang_etal_2019,calahan_etal_2020, zhang_etal_2021}. (7) Dust surface density at the planet's location \protect \citep{huang_etal_2018, nomura_etal_2021, fedele_etal_2017, benisty_etal_2021} accounting for an assumed factor of 7 uncertainty in opacity. Because most gaps are only marginally resolved in mm continuum emission we are likely overestimating $\Sigd$ (see section \ref{subsec:sigd}). Daggers ($\dagger$) mark the five gaps which \protect\citet{jennings_etal_2021} found to have lower dust continuum intensities and hence lower $\Sigd$ than \protect\citet{huang_etal_2018} found.
(8) Disc aspect ratio, mostly taken from \protect\citet{zhang_etal_2018} except for those marked with an asterisk (see section \ref{subsec:hr}).
The values for $t_{\rm age}$, $M_{\rm p}$, $\Sigg$, $\Sigd$, and $h/r$ are uncertain and quoted to only one significant figure.}
  \label{tab:data}
\end{table*}

\section{Observational Data}
\label{sec:data}
Table \ref{tab:data} compiles disc and candidate planet properties taken directly from the literature or derived therefrom. Quantities of interest include the host stellar mass $M_\star$ and planet mass $M_{\rm p}$,
the gas and dust surface densities $\Sigg$ and $\Sigd$ at planet orbital distance $r$, and the local disc aspect ratio $h/r$. To the nineteen DSHARP protoplanets modeled by \cite{zhang_etal_2018} we add the one candidate protoplanet in HD 169142 \citep{dong_fung_2017}, the two candidates in TW Hya \citep{dong_fung_2017}, and the two confirmed planets in PDS 70 \citep{keppler_etal_2018, haffert_etal_2019}.
%To the nineteen DSHARP protoplanets modeled by \cite{zhang_etal_2018} we add the candidate protoplanet in HD 169142 \citep{dong_fung_2017, fedele_etal_2017} and the two confirmed planets in PDS 70 \citep{keppler_etal_2018, haffert_etal_2019}.

\subsection{Gas surface densities $\Sigg$}
\label{subsec:gsd}
We focus on systems for which emission from optically thin C$^{18}$O has been spatially resolved. More abundant isotopologues of CO are usually optically thick (including $^{13}$CO; \citealt{vdm_etal_2015, favre_etal_2019}) and as such yield only lower limits on $\Sigg$. Hydrogen surface densities derived from C$^{18}$O emission are nonetheless uncertain. The gas-phase CO:H$_2$ abundance ratio $X_{\rm CO}$ is hard to determine; it may be lower in protoplanetary discs than in the interstellar medium (ISM) because of condensation of CO onto dust grains or 
chemistry driven by cosmic rays
(e.g. \citealt{schwarz_etal_2018}).
We will try to account for the possibility of gas-phase CO depletion in what follows. 
Surface densities at the centers of
gaps where planets supposedly reside
may also be overestimated if gaps are spatially underresolved, with emission from gap edges contaminating the emission at gap center. Such contamination might not be too serious for the gaps considered in this subsection, whose radial widths are larger (marginally) than the corresponding ALMA beams. Also, simulations predict gas density variations less than a factor of 10 across gaps for $M_{\rm p} \lesssim 60 \, M_\oplus$ (e.g. \citealt{dong_etal_2017}, top left panel of their figure 8).   

Of the twenty-four planets in our sample we have measurements of $\Sigg$ based on C$^{18}$O emission  (in conjunction with other emission lines) 
for nine 
planets in six systems.

\subsubsection{Gas surface density: AS 209}
\label{subsubsec:sigg_as209}

The protoplanetary disc around AS 209 has five annular gaps in dust continuum emission at orbital radii $r \approx 9,\,24,\,35,\,61,$ and $99$ au \citep{guzman_etal_2018, huang_etal_2018}. \cite{zhang_etal_2018} demonstrated that the four gaps from $24-99$ au can be produced by a single planet situated within the outermost gap at 99 au (see their figure 19). \cite{favre_etal_2019} and \cite{zhang_etal_2021} observed a broad gap in C$^{18}$O (2-1) emission extending from $\sim$60 to 100 au. 
\cite{favre_etal_2019} modeled the emission using the DALI thermo-chemical code \citep{bruderer_2013} and took the gas density profile to include two gaps coincident with the dust gaps at 61 and 99 au. They showed these two gas gaps could be smeared by the ALMA beam into a single wide gap like that observed in molecular emission. At the bottom of the 99 au gap where the planet is believed to reside, they fitted $\Sigg \approx 0.04-0.08\,\gcm$ (see the blue and green curves in their figure 6).

\cite{zhang_etal_2021} modeled the  C$^{18}$O emission using the thermo-chemical code RAC2D \citep{du_bergin_2014}, including only a single gap.
They initialized their models with an ISM-like $X_{\rm CO}$ which decreased from CO freeze-out and chemistry over the simulation duration of 1 Myr. They found $\Sigg \approx 0.08 \,\gcm$ at 99 au (their figure 16), similar to \cite{favre_etal_2019}. \cite{zhang_etal_2021} also considered the possibility that there is no gap in H$_2$ and that the observed gap in C$^{18}$O is due to $X_{\rm CO}$ being somehow locally depleted beyond the predictions of RAC2D. Assuming a total gas-to-dust ratio of $10$ by mass (their table 2), they estimated in this super-depleted $X_{\rm CO}$ scenario $\Sigg \approx 0.4 \,\gcm$ at 99 au (their figure 5; see also \citealt{alarcon_etal_2021} who advocated on separate grounds for a depleted CO:H$_2$ ratio in AS 209).

We combine the results of \cite{favre_etal_2019} and \cite{zhang_etal_2021} to consider $\Sigg = 0.04-0.4 \,\gcm$ for the gap AS 209-G2 at $r = 99$ au (Table \ref{tab:data}).

\subsubsection{Gas surface density: HD 163296}
\label{subsubsec:sigg_isella}

Dust continuum observations of HD 163296 reveal gaps centred at 10, 48, and 86 au \citep{isella_etal_2016, huang_etal_2018}, each opened by a separate planet according to \cite{zhang_etal_2018}. 
\cite{isella_etal_2016} fitted the emission at $r \gtrsim 25$ au from three CO isotopologues
assuming an ISM-like value of $X_{\rm CO}$ and found $\Sigg \approx 2.5-10\,\gcm$ at 48 au and $\Sigg \approx 0.1-2 \,\gcm$ at 86 au (their figure 2, blue curve; note the gap radii in their figure are about 20\% larger than the values we use because they used a pre-GAIA source distance). The thermo-chemical models of \cite{zhang_etal_2021} predict roughly consistent values of $\Sigg \approx 1 \,\gcm$ and $\Sigg \approx 0.6 \,\gcm$ respectively (see their figure 16). If instead the CO:H$_2$ abundance is much lower than predicted by their models and the gas-to-dust ratio is $60$ (their table 2), then  $\Sigg \approx 40\,\gcm$ and $\Sigg \approx \, 20\, \gcm$ (their figure 5).

The above estimates of $\Sigg$ are combined to yield the full ranges quoted in Table \ref{tab:data}, $\Sigg = 1-40\,\gcm$ for HD 163296-G2 at 48 au, and $\Sigg = 0.1-20\,\gcm$ for HD 163296-G3 at 86 au.

\subsubsection{Gas surface density: IM Lup}
\label{subsubsec:sigg_IMLup}
IM Lup presents a shallow gap at $r \approx 117$ au in millimeter continuum images \citep{huang_etal_2018} which \cite{zhang_etal_2018} attributed to a 10-30 $\Mearth$ planet.
\cite{zhang_etal_2021} found $\Sigg \approx 0.1 \,\gcm$ in their RAC2D modeling which includes CO freeze-out and chemistry (their figure 16). Alternatively, for a super-depleted CO:H$_2$ scenario and gas-to-dust ratio of 100 (their table 2), $\Sigg \approx 10\,\gcm$ (their figure 5). We allow for both possibilities and take $\Sigg = 0.1-10\,\gcm$.

\subsubsection{Gas surface density: HD 169142}
\label{subsubsec:sigg_fedele}
HD 169142 features an annular gap at 50 au in near-infrared scattered light \citep{quanz_etal_2013} and mm-wave continuum emission \citep{fedele_etal_2017}. In C$^{18}$O emission, \citet{fedele_etal_2017} 
found evidence for a cavity inside $r \approx 50$ au and used DALI to infer $\Sigg \approx 0.1-0.2\,\gcm$ near the cavity outer edge (their figure 5, top left panel, blue curve). 
Midplane gas temperatures according to the same thermo-chemical model (figure 5, bottom center panel) are $\sim$$30-40$ K, higher than the CO freeze-out temperature of $\sim$20 K. CO may still be depleted for other reasons  (\citealt{schwarz_etal_2018})
but without a model we cannot quantify the uncertainty.

\subsubsection{Gas surface density: TW Hya}
Near-infrared scattered light images reveal annular gaps at $r\approx 21$ au (TW Hya-G1) and 85 au (TW Hya-G2)  \citep{van-boekel_etal_2017}. The inner gap may also have been detected in the mm continuum by \cite{andrews_etal_2016}, albeit slightly outside the gap in scattered light (their figure 2). \cite{nomura_etal_2021} studied TW Hya in the $J=3-2$ transition of C$^{18}$O. Near the inner gap they measured a wavelength-integrated line flux of 5 mJy beam$^{-1}$ km s$^{-1}$ (their figure 2, middle panel, orange curve). We extrapolate their radial intensity profile, which cuts off at 80 au, to the outer gap at 85 au and estimate there a flux of 1 mJy beam$^{-1}$ km s$^{-1}$. 
Using temperatures of 60 K and 20 K for the inner and outer gaps respectively (section \ref{subsubsec:hr_tw}), we convert these fluxes to C$^{18}$O column densities \citep[e.g.][their equation B2]{mangum_shirley_2015}. Then assuming $X_{\rm CO} = 10^{-4}$ and a $^{18}$O$:^{16}$O gas-phase abundance ratio of $2 \times 10^{-3}$ \citep[e.g.][]{qi_etal_2011}, we calculate $\Sigg = 0.04\,\gcm$ and $0.008\,\gcm$ for the two gaps.
These surface densities could be underestimated by factors of 70 and 30 (assuming a gas-to-dust ratio of 100), respectively, if CO is somehow super-depleted  (\citealt{zhang_etal_2019}, their figure 8, second panel, blue curve). Thus we have $\Sigg \approx 0.04-2.8\,\gcm$ for TW Hya-G1 and $\Sigg \approx 0.008 - 0.24\,\gcm$ for TW Hya-G2.

\subsubsection{Gas surface density: PDS 70}
\label{subsubsec:gsd_pds70}

\cite{facchini_etal_2021} studied PDS 70 in the $J=2-1$ transition of $\rm C^{18}O$, but the beam size at this wavelength was too large to resolve the cavity within which the two planets reside. We therefore resort to using the C$^{18}$O emission just outside the cavity to place an upper limit on the gas surface density near the planets. Following the same procedure as for TW Hya, we convert the observed C$^{18}$O flux outside the cavity into an H$_2$ surface density. 
We use 30 K as the gas excitation temperature at 80 au, close to the H$_2$CO excitation temperature measured by \citet[][their figure 7]{facchini_etal_2021}. The gas 
surface density so derived is $\Sigg = 8 \times 10^{-2}\,\gcm$.
We pair this upper limit with a lower limit derived from optically thick $^{12}$CO in the $J=3-2$ transition. The smaller beam associated with the higher frequency of this transition allowed the cavity to be resolved by \citet{facchini_etal_2021}. Near the positions of the planets at $r \sim 22-34$ au, the observed $^{12}$CO line flux is 37 K km s$^{-1}$, which for an assumed temperature of $T \sim 60$ K (scaled from 30 K at 80 au using $T \propto r^{-3/7}$ for a passively irradiated disk; \citealt{chiang_goldreich_1997}) yields a lower bound on $\Sigg$ of $8\times 10^{-4}\,\gcm$. The temperatures quoted above are high enough to avoid CO freeze-out, though we cannot rule out that gas-phase CO is depleted from other effects \citep{schwarz_etal_2018}.

\subsection{Dust surface densities $\Sigd$}
\label{subsec:sigd}
The dust surface density $\Sigd$ and the dust optical depth $\tau$ (typically evaluated at mm wavelengths) are related by
\begin{equation}
    \Sigd = \frac{\tau}{\kappa} \label{eqn:Sigd}.
\end{equation}
The opacity $\kappa$ depends on the grain size distribution. We allow $\kappa$ to range from 0.4 to 3 cm$^{2}$ per gram of dust,\footnote{The ALMA continuum data used in our study were taken at wavelengths ranging from 0.85 mm to 1.27 mm. Our adopted factor-of-seven uncertainty in $\kappa$ is larger than the expected variation of $\kappa$ across this wavelength range (at fixed grain size), so for simplicity we ignore the wavelength dependence on $\kappa$ when computing $\Sigd$ from the ALMA observations.} following model size distributions by \citet{birnstiel_etal_2018} which include grains up to 1 cm in size. For a given temperature $T$, we infer $\tau$ from the specific intensity $I_{\nu}$:
\begin{equation}
    I_{\nu} = B_{\nu}(T)\left(1 - e^{-\tau}\right)
\label{eqn:Inu}
\end{equation}
where $B_{\nu}(T)$ is the Planck function.

Surface densities may be overestimated inside gaps that are spatially underresolved. This may be a much more serious problem for $\Sigd$ than for $\Sigg$ (cf. section \ref{subsec:gsd}) because simulations predict dust surface densities to vary by orders of magnitude
across gaps carved by $M_{\rm p} \gtrsim 10 \, M_\oplus$ planets, with dust gradients steepened by aerodynamic drag (e.g. figure 1 of \citealt{paardekooper_mellema_2004}; figure 8 of \citealt{dong_etal_2017}; figure 4 of \citealt{binkert_etal_2021}).
We caution that the values of $\Sigd$ we derive may therefore be grossly overestimated. See also \cite{jennings_etal_2021}, whose work we describe below.

\subsubsection{Dust surface density: DSHARP systems}
For all nineteen of the DSHARP planets, we estimate $\Sigd$ at gap center using equation (\ref{eqn:Sigd}) and the optical depth profiles in figure 6 of \cite{huang_etal_2018}.

Three of the DSHARP gaps, AS 209-G2 ($r = 99$ au), Elias 24 ($57$ au), and HD 143006-G1 ($22$ au), have bottoms that fall below the 2$\sigma$ noise floor at $\tau = 10^{-2}$, although apparently only marginally so. For Elias 24 and HD 143006-G1, we set $\tau = 10^{-2}$. For AS 209-G2, while portions of the gap lie below the noise floor, at gap center there is actually a local maximum (ring) of emission that rises above the noise. We use the peak optical depth of this ring, $\tau \approx 2 \times 10^{-2}$, to evaluate $\Sigd$. The ``W"-shaped gap profile
for AS 209-G2 resembles transient structures found in two-fluid simulations of low-mass planets embedded in low-viscosity discs \citep[][]{dong_etal_2017, zhang_etal_2018}.

In their re-analysis at higher spatial resolution of the DSHARP data, \citet[][their figure 12]{jennings_etal_2021} found evidence for deeper dust gaps than originally measured by \cite{huang_etal_2018} for GW Lup, HD 163296-G2, SR 4, AS 209-G1, and HD 143006-G1. The re-analyzed gaps in the mm continuum are deeper by at least a factor of 10 and in most cases much more. We make a note
of these sources in Table \ref{tab:data},
but continue to use the \citet{huang_etal_2018} results to compute $\Sigd$ for consistency with the rest of our analysis.

\subsubsection{Dust surface density: TW Hya}
\label{subsubsec:twhya}
For TW Hya-G1 at 21 au we calculate $\Sigd = %0.054-0.38\,\gcm$ 
0.06-0.4\,\gcm$ using equations (\ref{eqn:Sigd})-(\ref{eqn:Inu}) with $I_{\nu} = 10$ mJy beam$^{-1}$ (figure 2, middle panel, purple curve of \citealt{nomura_etal_2021}) and $T = 60$ K (section \ref{subsubsec:hr_tw}). For TW Hya-G2 at 85 au we estimate $\Sigd = 0.002-0.02\,\gcm$ using
$I_{\nu} \sim 10^{-1}\,$ mJy beam$^{-1}$ (this is an extrapolation of the intensity profile measured by \citealt{nomura_etal_2021} which does not extend beyond $\sim$70 au) and $T = 20$ K. Our estimates of $\Sigd$ are consistent with those plotted in figure 3 of \cite{nomura_etal_2021}.  

\subsubsection{Dust surface density: HD 169142 and PDS 70}
\label{subsubsec:dsd_hd169142_pds70}
For the gap in HD 169142 and central clearing in PDS 70, the continuum emission falls below the 3$\sigma$ noise floor of the observations (0.21 mJy beam$^{-1}$ and 0.0264 mJy beam$^{-1}$, respectively; 
\citealt{fedele_etal_2017, benisty_etal_2021}). For assumed temperatures of $T=40$ K in HD 169142 and $T = 60$ K in PDS 70, the corresponding upper limits on the dust column 
are $\Sigd \leq 10^{-2}\,\gcm$ and $\Sigd \leq 2 \times 10^{-2}\,\gcm$, respectively.

\subsection{Disc scale height $h$}
\label{subsec:hr}

\subsubsection{Disc scale height: DSHARP systems}
\label{subsubsec:hr_dsharp}
For most of the DSHARP sample, we adopt the $h/r$ estimates listed in table 3 of \cite{zhang_etal_2018}, computed assuming a passive disc irradiated by its host star. For AS 209-G2, all three gaps in HD 163296, and IM Lup, we use more sophisticated 
estimates of $h/r$ from table 2 of \citet[][]{zhang_etal_2021}. In AS 209, \cite{zhang_etal_2021}
used the observed $^{13}$CO emission surface to fit for $h/r = 0.06(r/100\,\rm au)^{0.25}$ (since their CO isotopologue data do not extend down to AS 209-G1 at 9 au, we only apply this scale height relation to AS 209-G2 at 99 au). Note that
\cite{zhang_etal_2018} found a similar value, $h/r = 0.05$ at 99 au, reproduced the multi-gap structure of AS 209. For HD 163296 and IM Lup, \cite{zhang_etal_2021} obtained excellent fits to the spectral energy distributions (their figure 4) using $h/r = 0.084(r/100\,\rm au)^{0.08}$ and $h/r = 0.1(r/100\,\rm au)^{0.17}$, respectively.\footnote{
The values of $h$ 
obtained by \cite{zhang_etal_2021},
if interpreted as a standard isothermal 
hydrostatic disc gas scale height,
imply gas temperatures that differ from
the gas temperatures obtained from thermo-chemical and radiative transfer modeling by these same authors (their figure 7). This is because the latter temperatures are solved assuming a fixed density field (specified by $h$), and no provision is made to iterate the density field to ensure simultaneous hydrostatic and radiative equilibrium. There is no rigorous justification for using either set of temperatures to compute the gas scale height; we use $h$ as tabulated
by \cite{zhang_etal_2021} for convenience (their parameters $H_{100}$ and $\psi$ in table 2, inserted into their equation 3 with $\chi = 1$).
}

\subsubsection{Disc scale height: HD 169142}
\cite{fedele_etal_2017} simultaneously fitted the surface density and temperature profiles 
in HD 169142. We use their 
aspect ratio of $h/r = 0.07$ at the gap position; this is
similar to the value of $h/r = 0.08$ found by \cite{dong_fung_2017} in fitting the gap width to their planet-disk simulations. The aspect ratio relates to the disc midplane temperature following $h/r = c_s/(\Omega r)$ where $c_s = \sqrt{kT/\mu m_{\rm p}}$ is the sound speed, $\mu = 2.4$ is the mean molecular weight, $k$ is Boltzmann's constant, $m_{\rm p}$ is the proton mass, $\Omega = \sqrt{GM_{\star}/r^3}$ is the Keplerian orbital frequency, and $G$ is the gravitational constant. The implied temperature at the gap position is 40 K.

\subsubsection{Disc scale height: TW Hya}
\label{subsubsec:hr_tw}
\cite{calahan_etal_2020} obtained a good fit to the spectral energy of TW Hya using $h/r = 0.11 \,(r/400\,\rm au)^{0.1}$ (their figure 5). From this scaling we calculate $h/r = 0.08$ and $h/r = 0.09$ at the inner ($r= 21$ au) and outer ($r= 85$ au) gap positions, respectively. These aspect ratios correspond to temperatures of 58 K and 20 K.

\subsubsection{Disc scale height: PDS 70}
Absent direct estimates of the temperature in the PDS 70 cavity, we estimate the disc scale height using arguments similar to those in section \ref{subsubsec:gsd_pds70}. \cite{facchini_etal_2021} derived an H$_2$CO excitation temperature near the disc midplane of 33 K at 80 au (see their figure 7). Setting this equal to the gas kinetic temperature there, we calculate $h = c_s/\Omega$ at 80 au, and then scale to the location of the planets assuming $h/r \propto r^{2/7}$ \citep{chiang_goldreich_1997}. We find $h/r = 0.07$ and 0.08 at the locations of PDS 70b and c. \citet{keppler_etal_2018} found a similar temperature profile in their radiative transfer models of the PDS 70 disc.

\subsection{Planet and stellar masses}

\subsubsection{Planet masses $M_{\rm p}$ and stellar masses $M_{\star}$: DSHARP systems}
\label{subsubsec:masses_dsharp}
\cite{zhang_etal_2018} used their grid of planet-disk hydrodynamical simulations to
estimate the planet masses 
that can reproduce the 
radial widths and in some cases the
radial positions of 
DSHARP gaps. Their table 3 gives $M_{\rm p}$ for different 
dust grain size distributions and the \cite{shakura_sunyaev_1973} viscosity parameter $\alpha$. Here we consider $\alpha = 10^{-5}-10^{-3}$. 
Values of $\alpha \gtrsim 10^{-2}$ appear incompatible with various mm-wave observations \citep{pinte_etal_2016, teague_etal_2016, flaherty_etal_2017} including the multi-gap structure of some discs \citep{dong_etal_2017, zhang_etal_2018}. Planet masses inferred by \citet{zhang_etal_2018} increase with viscosity as $M_{\rm p} \propto \alpha^{1/3}$; more massive planets are required to counteract faster diffusion of gas to open the same width gap. 
Thus to obtain an upper limit on $M_{\rm p}$ we use the entries corresponding to $\alpha = 10^{-3}$ in table 3 of \cite{zhang_etal_2018}, and for a lower limit we take their $M_{\rm p}$ values for $\alpha = 10^{-4}$ and divide by two in a crude extrapolation down to $\alpha = 10^{-5}$. This procedure gives a range on $M_{\rm p}$ that typically spans a factor of 8, and that encompasses the variation in $M_{\rm p}$ due to different dust size distributions.

For HD 163296-G3, we increase the upper bound on $M_{\rm p}$ to $1 M_{\rm J}$ based on the fit to non-Keplerian kinematical data from \cite{teague_etal_2018} and \cite{teague_etal_2021}. Note that their mass estimate is based on planet-disc simulations assuming $\alpha = 10^{-3}$, and that lowering $\alpha$ lowers the planet mass required to reproduce the same non-Keplerian velocity signature ($M_{\rm p} \propto \alpha^{0.41}$ according to equation 16 of \citealt{zhang_etal_2018}).
For Elias 24, the $M_{\rm p}$ range in Table \ref{tab:data} overlaps with mass estimates inferred from infrared photometry of an unconfirmed point source at $r \approx 55$ au (\citealt{jorquera_etal_2021}). The candidate point source in HD 163296 is located at $r \approx 67$ au (\citealt{guidi_etal_2018})
and as such cannot be identified with the planets inferred at $r \approx 48$ and 86 au (\citealt{zhang_etal_2018}) of interest here.

For most of the DSHARP systems we use stellar masses derived from  stellar evolution tracks and listed in table 1 of \cite{andrews_etal_2018}. For IM Lup, AS 209, and HD 163296 we use instead $M_{\star} = 1.2,\,1.1,$ and $2.0\,\Msun$ respectively, calculated from gas disc rotation curves by \cite{teague_etal_2021} 
as part of the MAPS survey \citep{oberg_etal_2021}.

\subsubsection{Planet and stellar masses: HD 169142 and TW Hya}
By comparing the observed gaps in HD 169142 and TW Hya with their grid of planet-disk simulations, \cite{dong_fung_2017} estimated planet masses of $M_{\rm p} = 0.2-2\,\Mj$ in HD 169142, $0.05-0.5\,\Mj$ in TW Hya-G1, and $0.03-0.3\,\,\Mj$ in TW Hya-G2. These values were derived for viscosities $\alpha = 10^{-4}-10^{-2}$. For consistency with our adopted range of $\alpha = 10^{-5}-10^{-3}$ (see section \ref{subsubsec:masses_dsharp}), we apply the same $M_{\rm p} \propto \alpha^{1/3}$ scaling from \cite{zhang_etal_2018} and shift all limits from \cite{dong_fung_2017} down by a factor of two.

The stellar masses of HD 169142 and TW Hya are $M_{\star} = 1.65 \, \Msun$ and $M_{\star} = 0.8\,\Msun$, respectively \citep{blondel_dije_2006, ducourant_etal_2015}.

\subsubsection{Planet and stellar masses: PDS 70} 
Masses of PDS 70b and c derived from near-infrared spectral energy distributions and cooling models
vary widely \citep{muller_etal_2018, keppler_etal_2018, christiaens_etal_2019, mesa_etal_2019, wang_etal_2020, stolker_etal_2020}. For simplicity we take $M_{\rm p}= 1-10 \,\Mj$ for both planets, which brackets most literature estimates 
relying on cooling models.\footnote{In later sections, and in particular the discussion surrounding Figures \ref{fig:sed2} and \ref{fig:sed}, we will consider the possibility that the observed near-infrared emission does not 
arise from a planet passively cooling into empty space (as assumed by the cooling models) 
but one which sits below a hot accretion shock boundary layer. In this scenario our adoption of planetary masses for PDS 70b and c drawn from the passive cooling models will lead to an inconsistency.  
Our hope is that our estimated mass range of $1-10 \,M_{\rm J}$ is not seriously impacted; it should not be as long as the posited accretion luminosities are not much larger than the observed near-infrared luminosities ($10^{-4} L_\odot$;
\citealt{wang_etal_2020}). It may also help to have the accretion columns be localized to ``hot spots'' on the planet, like they are for magnetized T Tauri stars (e.g. \citealt{gregory_etal_2006}), leaving the rest of the planet passively cooling.
} This range is consistent with masses for PDS 70b derived by \cite{wang_etal_2021} based on dynamical stability constraints. They calculated a 95\% upper limit of 10 $M_{\rm J}$, with roughly equal probability per log mass from 1-10 $M_{\rm J}$ (see their figure 5). 

The mass of the star is $M_{\star} = 0.88 \, \Msun$ \citep{keppler_etal_2019}.

\begin{figure*} 

\includegraphics[width=0.95\textwidth]{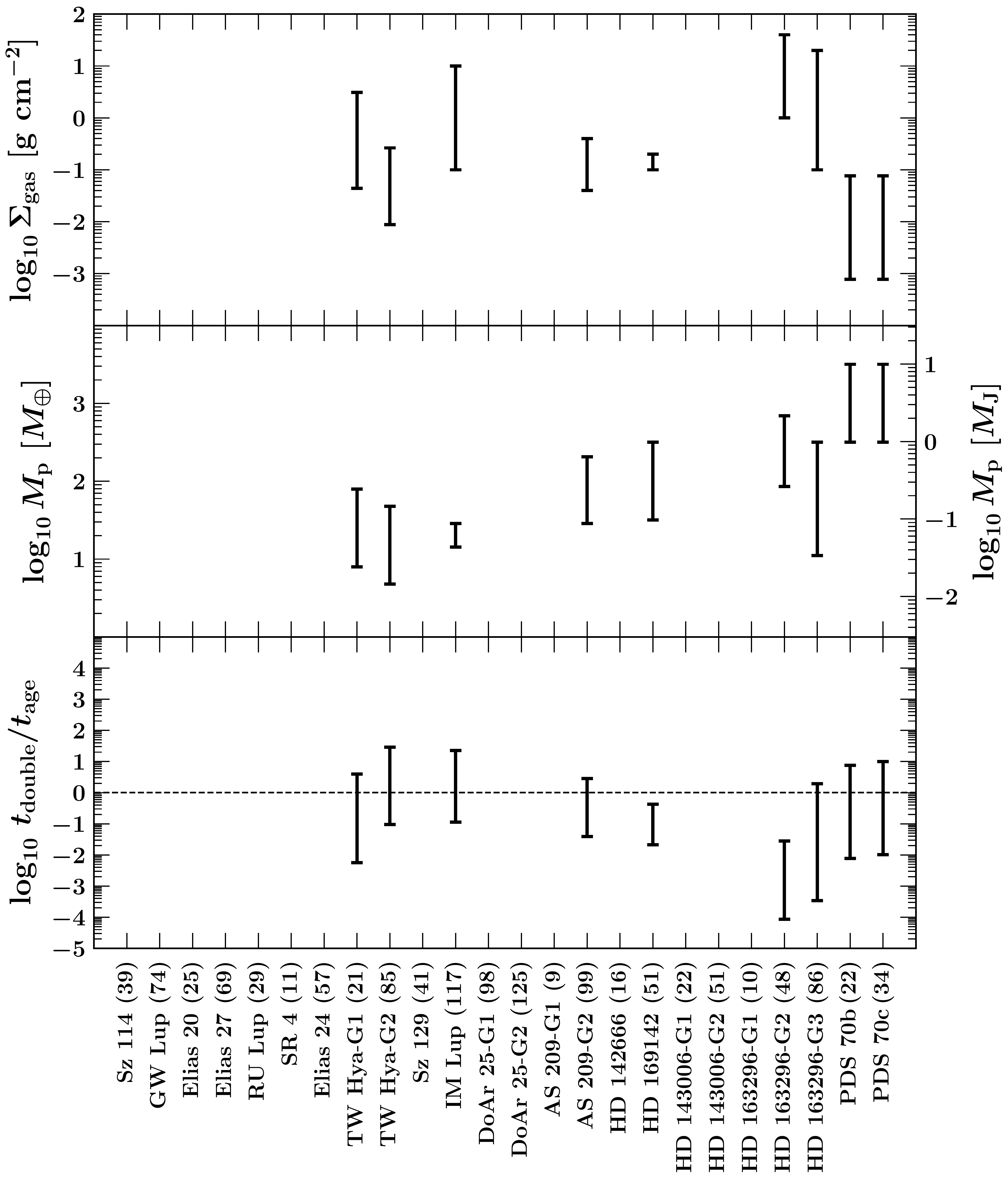}
\vspace{-2mm}
\caption{{\textbf{Top panel:}} 
Gas surface density $\Sigg$ at the centers of gaps hypothesized (confirmed for PDS 70) to host planets, for the subset of systems detected in C$^{18}$O.
System names on the 
horizontal axis are followed by 
``G\#'' to identify a specific gap in systems with multiple gaps, and by the planet's orbital radius $r$ in units of au in parentheses. Systems are listed in order of increasing host stellar mass $M_{\star}$, except for PDS 70 which as the only transitional disc in our sample is placed at the end. For TW Hya we estimate $\Sigg$ from the C$^{18}$O line flux \protect \citep{nomura_etal_2021},
while for AS 209, IM Lup, HD 169142, and HD 163296 $\Sigg$ derives from detailed thermo-chemical modeling  \protect \citep{favre_etal_2019, zhang_etal_2021,  fedele_etal_2017, isella_etal_2016}. For PDS 70, we bound $\Sigg$ using the C$^{18}$O 2-1 flux outside the disc cavity and the $^{12}$CO 3-2 flux inside the cavity \protect \citep{facchini_etal_2021}. {\textbf{Middle panel:}} Planet masses $M_{\rm p}$. For the non-PDS 70 planets $M_{\rm p}$ is drawn from gap-fitting simulations by \protect \cite{zhang_etal_2018} and \protect \cite{dong_fung_2017} scaled to  viscosities $\alpha = 10^{-5} - 10^{-3}$. For PDS 70b and c, $M_{\rm p} = 1-10\,\Mj$  roughly encompasses values from cooling models \protect \citep[e.g.][]{haffert_etal_2019}.
{\textbf{Bottom panel:}} Mass doubling times $t_{\rm double} \equiv M_{\rm p}/\dot{M}_{\rm p}$ calculated assuming accretion at the Bondi rate (equation \ref{eqn:Mdot_bondi}). Accretion rates $\dot{M}_{\rm p}$ are evaluated using $\Sigg$ and $M_{\rm p}$ as above, plus $h$ and $r$ compiled in Table \ref{tab:data}.
Error bars account for uncertainties in $\Sigg$ and $M_{\rm p}$, with $t_{\rm double} \propto 1/(M_{\rm p}\Sigg)$. Eight of the nine hypothesized planets may have $t_{\rm double}$ comparable to their system age $t_{\rm age}$, a plausible result that if true suggests these planets are undergoing their last doublings. }
  \label{fig:CO}
\end{figure*}

\begin{figure*} 
\includegraphics[width=\textwidth]{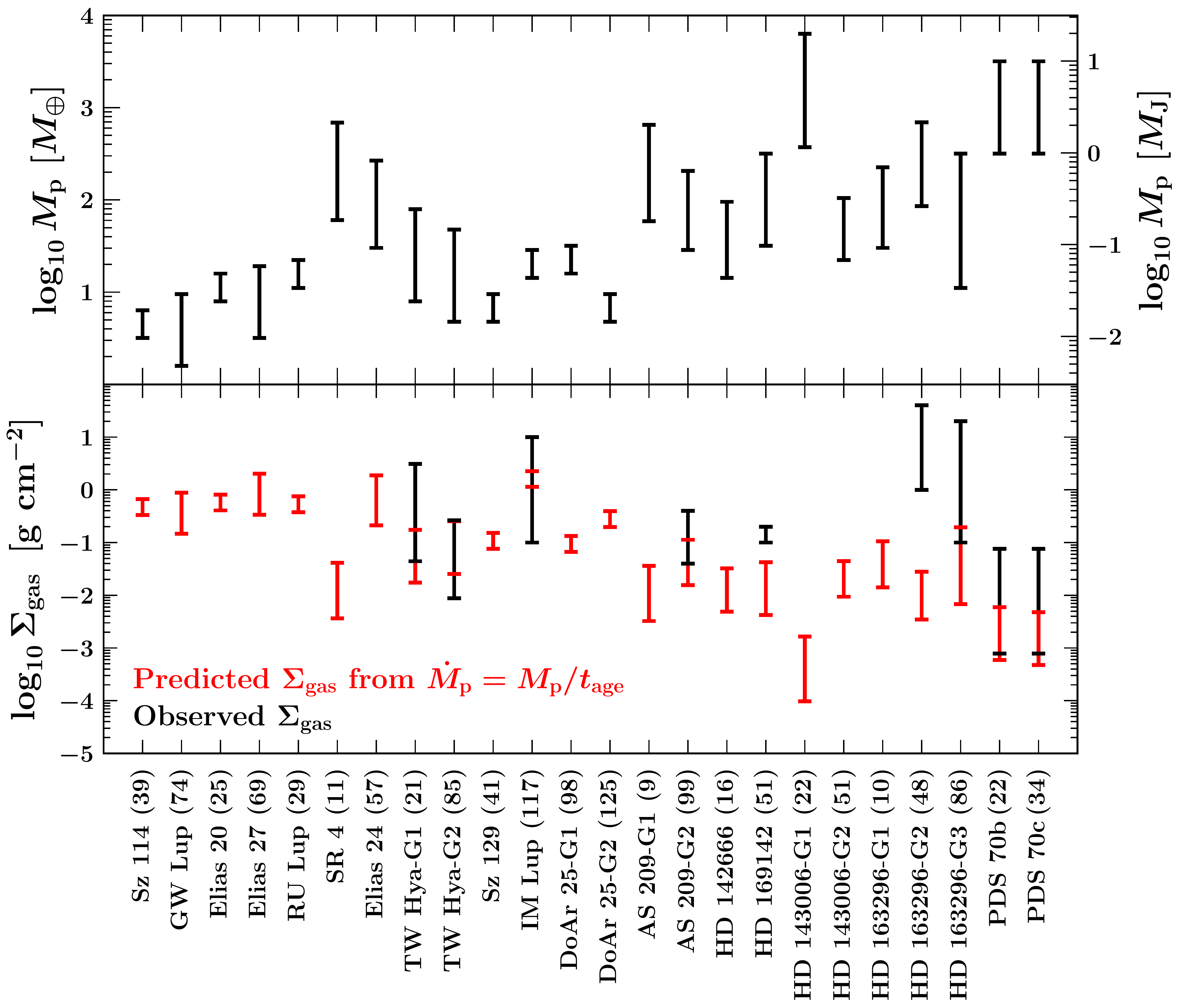}
\vspace{-2mm}
\caption{
\textbf{Top panel:} Same as the middle panel of Figure \ref{fig:CO}, with $M_{\rm p}$ now plotted for all 24 planets in our sample. \textbf{Bottom panel:}
Same as the top panel of Figure \ref{fig:CO}, with red points added to mark $\Sigg$ predicted from equation (\ref{eqn:backward}). The prediction assumes that each planet is accreting at the Bondi-like rate given by (\ref{eqn:Mdot_bondi}) and that $\dot{M}_{\rm p} = M_{\rm p}/t_{\rm age}$ (equivalently $t_{\rm double} = t_{\rm age}$; for Elias 24 we use the upper limit on $t_{\rm age}$). The last equality is motivated by the bottom panel of Figure \ref{fig:CO}. The evaluation of (\ref{eqn:backward}) uses $M_{\rm p}$ in the top panel plus $r$ and $h$ listed in Table \ref{tab:data}. Red error bars reflect the range of $M_{\rm p}$ values considered.}
  \label{fig:bondi_prediction_gas}
\end{figure*}

\begin{figure*} 
\includegraphics[width=\textwidth]{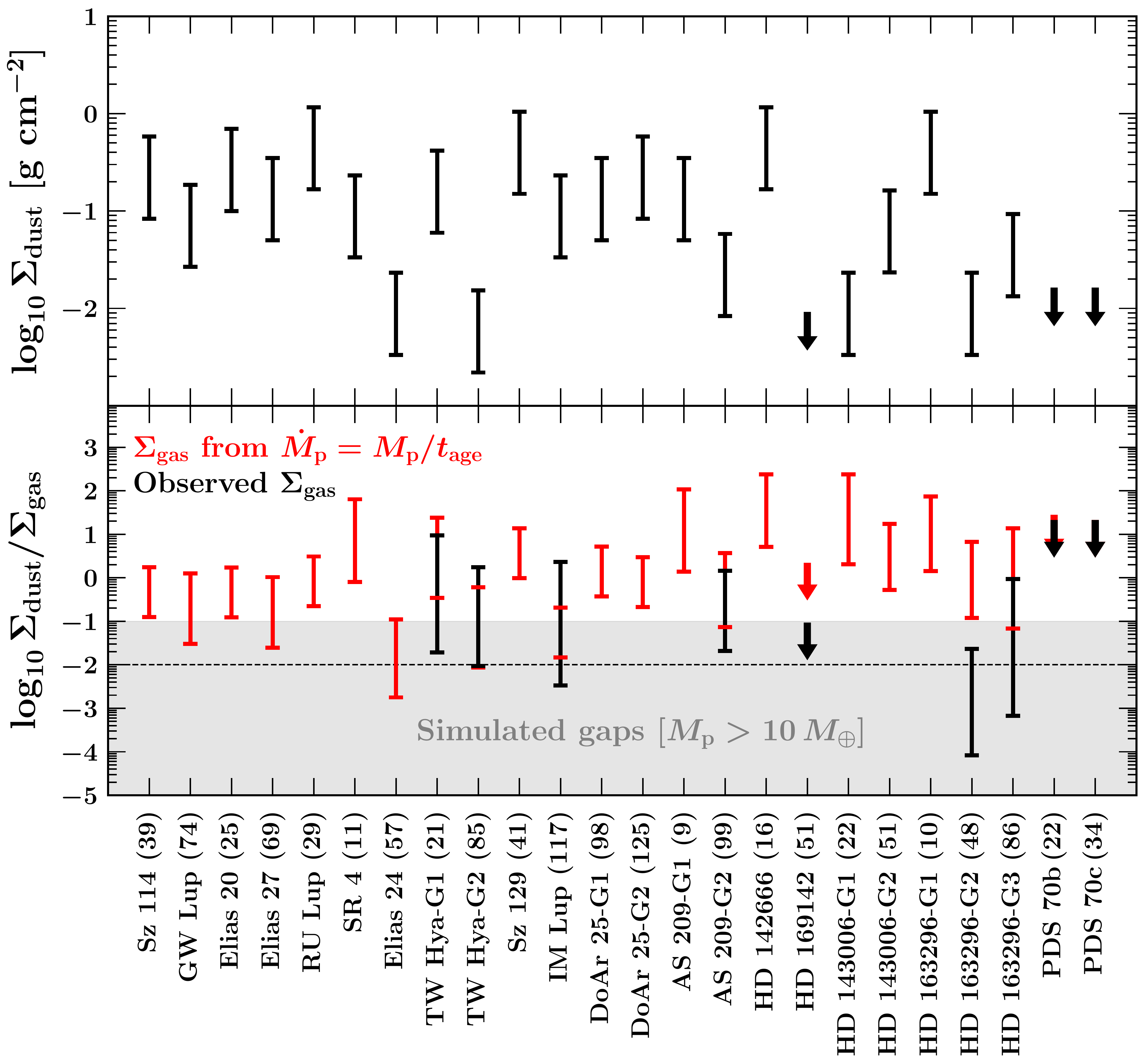}
\vspace{-2mm}
\caption{\textbf{Top panel:} Same as the top panel of Figure \ref{fig:CO} but plotting
  dust surface densities $\Sigd$ inside gaps based on
  observed mm continuum intensities
  (\citealt{huang_etal_2018}), assuming the dust is optically thin. Error bars account for a factor of seven
  uncertainty in the grain opacity but do not account for the limited
  spatial resolution of the observations which may lead to $\Sigd$
  being overestimated. For HD 169142 and PDS 70, we plot upper limits
  on $\Sigd$ set by the noise floor of the observations \protect
  \citep{fedele_etal_2017, benisty_etal_2021}. \textbf{Bottom panel:} Gap
  dust-to-gas ratios $\Sigd/\Sigg$. In black we combine the values of $\Sigd$ 
  with $\Sigg$ inferred from C$^{18}$O intensity maps (Figure
  \ref{fig:CO}, top panel). In red we use the same
  $\Sigd$ data but take $\Sigg$ from equation (\ref{eqn:backward}),
  which presumes that planets accrete in a Bondi-like way with growth
  timescales $M_{\rm p}/\dot{M}_{\rm p}$ equal to system ages (Figure \ref{fig:bondi_prediction_gas}, red points). 
The shaded bar covers the range of dust-to-gas ratios found inside
gaps opened by planets with $M_{\rm p} > 10 M_\oplus$ in two-fluid
simulations by \protect \citet{dong_etal_2017}. Many of the empirical data
lie outside the simulated range, possibly because
dust gaps are still underresolved by ALMA and therefore $\Sigd$ is overestimated. In support of this idea, \protect \citet{jennings_etal_2021} re-analyzed the DSHARP data and found deeper bottoms to the dust gaps in GW Lup, SR 4, AS 209-G1, HD 143006-G1, and HD 163296-G2.}
  \label{fig:dust}
\end{figure*}

\section{Planetary Accretion Rates and \\Gap Properties}
\label{sec:results}
Most of the gap-opening planets in our sample 
have inferred masses large enough 
that the self-gravity of their
gas envelopes should be significant.
In this regime, further accretion of gas 
from the disc may no longer be limited by
Kelvin-Helmholtz cooling of the
envelope but may instead be regulated by the hydrodynamics
of the surrounding nebula. 
Hydrodynamical forces include, in principle, pressure, planetary gravity, stellar tides, and rotation, all in 3D. The simplest 
form of hydrodynamical accretion is Bondi accretion,
which describes spherically symmetric 
(zero angular momentum) flow onto a point mass  \citep[e.g.][]{frank_king_raine}. For a point mass
embedded in a disc, the Bondi rate is given by:
\begin{align}
    \dot{M}_{\rm p} = 0.25 \Omega r^2 \left(\frac{M_{\rm p}}{\Mstar}\right)^2\left(\frac{h}{r}\right)^{-4}\Sigg %\,.
    \label{eqn:Mdot_bondi}
\end{align}
(e.g.~\citealt{ginzburg_chiang_2019a}). 
While it ignores many effects, Bondi accretion appears to fit
the results of 3D isothermal hydrodynamic simulations
of accreting planets
by \cite{dangelo_etal_2003},
both in magnitude and in the $M_{\rm p}^2$
scaling (see figure 1 of \citealt{tanigawa_tanaka_2016} which re-prints their data). We have inserted in equation (\ref{eqn:Mdot_bondi}) a factor of 0.25 to match the normalization of the $\dot{M}_{\rm p} \propto M_{\rm p}^2$
relation measured by \cite{dangelo_etal_2003} at $M_{\rm p} \lesssim 30 \,M_\oplus$.
These authors found that at higher masses, $\dot{M}_{\rm p} (M_{\rm p})$ flattens away from an $M_{\rm p}^2$ scaling; this flattening is due to the conflating effect of gap opening, i.e. $\Sigma_{\rm gas}$ in equation (\ref{eqn:Mdot_bondi}) decreases with $M_{\rm p}$ for masses $\gtrsim 30\, M_\oplus$ as deeper gaps are opened.
Note that the accretion rates measured by \citet{dangelo_etal_2003} may be sensitive to their mass removal prescription, which reduces the mass density inside the innermost $\sim$10\% of the planet's Hill sphere by some fraction every simulation timestep. This prescription effectively renders the planet a sink cell and 
maximizes $\dot{M}_{\rm p}$.

The corresponding planet mass doubling time is
\begin{align}
    t_{\rm double} &\equiv \frac{M_{\rm p}}{\dot{M}_{\rm p}}.
    \label{eqn:tdouble}
\end{align} 
We test the theory of planet
accretion (Bondi accretion) by
comparing the doubling time $t_{\rm double}$ to the age of the system $t_{\rm age}$.
We expect $t_{\rm double} \gtrsim t_{\rm age}$. Steady and ongoing growth over the planet's age corresponds to 
 $t_{\rm double}/t_{\rm age} \sim 1$
 (equivalently $\dot{M}_{\rm p} \sim M_{\rm p}/t_{\rm age}$). Alternatively, we may be observing
 the planet after it has largely finalized its mass, in which case $t_{\rm double} > t_{\rm age}$.
We would not expect to find
$t_{\rm double} \ll t_{\rm age}$
as a planet is unlikely to be
caught in a short-lived episode
of significant growth.

\subsection{Bondi accretion rates $\dot{M}_{\rm p}$ from C$^{18}$O, and $t_{\rm double}$ vs. $t_{\rm age}$}
\label{subsec:main_results}
Figure \ref{fig:CO} plots the ratio $t_{\rm double}/t_{\rm age}$ for each of the nine  protoplanets with $\Sigg$ measured from C$^{18}$O observations (section \ref{subsec:gsd}). 
We find that $t_{\rm double}/t_{\rm age}$ appears in the majority of cases to be consistent with unity, although the error bars, which reflect the combined uncertainty in $\Sigg$ and $M_{\rm p}$, are large. The result $t_{\rm double}/t_{\rm age} \sim 1$ is physically plausible and indicates that the planets are currently accreting at rates $\dot{M}_{\rm p}$ close to the values time-averaged over their system ages, $\dot{M}_{\rm p} \sim M_{\rm p}/t_{\rm age}$. For TW Hya-G1, AS 209-G2, HD 169142, and HD 163296-G3, $t_{\rm double}/t_{\rm age} \sim 1$ if $M_{\rm p} \sim 10-30\,\Mearth$ and $\Sigg \sim 0.1\,\gcm$, near the lower ends of their ranges in Table \ref{tab:data}. Lower planet masses imply lower disc viscosities ($\alpha \sim 10^{-5}$) insofar as $M_{\rm p} \propto \alpha^{1/3}$ (\citealt{zhang_etal_2018} and section \ref{subsubsec:masses_dsharp}). Lower gas surface densities argue against speculative super-depleted CO scenarios
and support thermo-chemical models (DALI, \citealt{bruderer_2013}; RAC2D, \citealt{du_bergin_2014}) that account for the usual ways that CO can be depleted, freeze-out and chemistry.

The gap HD 163296-G2 at $r=48$ au does not meet our expectation that $t_{\rm double}/t_{\rm age} \gtrsim 1$. For this case the computed value of $t_{\rm double}/t_{\rm age} \lesssim 10^{-1}$ is driven by the large Jupiter-scale mass that \citet{zhang_etal_2018} inferred from the observed deep and wide dust gap (\citealt{isella_etal_2016, huang_etal_2018}). Such a massive planet seems inconsistent with there being hardly any depression in the gas at this location (\citealt{zhang_etal_2021}, their figure 16 and section 4.3.2).  \cite{rodenkirch_etal_2021} modeled the HD 163296 disc and found that a Jovian-mass planet at 48 au would deplete the gas density by a factor of $\sim$10 for $\alpha = 2 \times 10^{-4}$ (their figure 5). We speculate that a planet much less massive than Jupiter might reproduce the dust and gas observations and bring $t_{\rm double}$ into closer agreement with $t_{\rm age}$, if such a lower-mass planet caused only modest gas pressure variations that led to stronger dust variations via aerodynamic drift \citep[][]{paardekooper_mellema_2004, dong_etal_2017, drazkowska_etal_2019, binkert_etal_2021}. Another possibility is that there is no planet at all inside the HD 163296-G2 gap, as a planet can create gaps not centered on its orbit; figure 8 of \citet{dong_etal_2018} explicitly models this scenario for HD 163296.

For PDS 70b and c, $t_{\rm double}/t_{\rm age} \sim 1$ implies $\dot{M}_{\rm p} \sim 6 \times 10^{-7}- 6 \times 10^{-6}\,\,M_{\rm J}\,\rm yr^{-1}$ assuming $M_{\rm p} = 3\,M_{\rm J}$. These values are factors of 10-100 higher than estimates based on the observed U-band and H$\alpha$ fluxes \citep[e.g.][]{haffert_etal_2019, zhou_etal_2021}. Section \ref{subsec:observables} discusses possible reasons why.

\subsection{$\Sigg$ and $\Sigd$}
\label{subsec:backward}

Fifteen of the twenty-four gaps in our sample lack published C$^{18}$O data, precluding us from estimating $\Sigg$ and by extension $\dot{M}_{\rm p}$. 
An alternate approach is to reverse the logic by first assuming 
that $\dot{M}_{\rm p} \sim M_{\rm p}/t_{\rm age}$ (i.e. assuming $t_{\rm double}/t_{\rm age} \sim 1$), inserting this accretion rate into equation (\ref{eqn:Mdot_bondi}), and solving for $\Sigg$:
\begin{equation}
\Sigg = \frac{4M_{\rm p}}{r^2} \frac{1}{\Omega t_{\rm age}} \left( \frac{M_\star}{M_{\rm p}} \right)^2 \left( \frac{h}{r} \right)^4  \,\,\,\,\,\,\,\,\,\,\,\,\,\,\, {\rm if\,no\,C}^{18}{\rm O \, data.}
\label{eqn:backward}
\end{equation}
In other words, we assume the planets without C$^{18}$O data are like the eight we identified in Figure \ref{fig:CO} with C$^{18}$O data showing evidence for $t_{\rm double} \sim t_{\rm age}$ --- we posit that all are actively accreting from their surrounding discs, and that none are being observed in a short-lived growth phase. We view the values of $\Sigg$ so computed as predictions for future observations.

Figure \ref{fig:bondi_prediction_gas} shows in red the gas surface densities predicted by equation (\ref{eqn:backward}). The values range from $10^{-4}\,\gcm$ to $1\,\gcm$, with a factor of 2-10 uncertainty in any given system stemming from the adopted range of planet masses $M_{\rm p}$. Note that if $t_{\rm double}/t_{\rm age} > 1$, the red points shift down. Plotted in black for comparison are the $\Sigg$ values inferred from the measured C$^{18}$O fluxes (same data plotted in the top panel of Figure \ref{fig:CO}). In some systems the $\Sigg$ predictions (red) overlap only with the lower end of values bracketed by the data (black) because $t_{\rm double}/t_{\rm age} = 1$ favors lower $\Sigg$, as mentioned at the beginning of section \ref{subsec:main_results}.

The top panel of Figure \ref{fig:dust} displays $\Sigd$ inferred
from the mm continuum observations (section \ref{subsec:sigd}). The
error bars account for a factor of 7 uncertainty in the grain opacity
but not the potential overestimation of $\Sigd$ from gaps being inadequately resolved (more discussion on this below). In the bottom panel, we combine these values of $\Sigd$ with $\Sigg$ 
inferred from C$^{18}$O data to plot, in black, the empirical
dust-to-gas ratios $\Sigd/\Sigg$. The red points are semi-theoretical as
they take $\Sigg$ from
Figure \ref{fig:bondi_prediction_gas} as predicted by equation
(\ref{eqn:backward}).

The shaded bar $\Sigd/\Sigg < 10^{-1}$ in Figure \ref{fig:dust} marks the range of dust-to-gas ratios found in simulated
gaps (\citealt{dong_etal_2017}). The simulations are initialized with spatially uniform
dust-to-gas ratios near the solar value of $\sim$10$^{-2}$. The ratios then evolve as aerodynamic drag causes dust to drift relative to gas.
Across most of the gap, dust-to-gas ratios decrease as dust is diverted
into local gas pressure maxima near gap edges. One gas pressure maximum coincides with the planet's orbit at gap center. Dust that collects in this co-orbital gas ring raises $\Sigd/\Sigg$ to
$\sim$0.1 (about ten times the solar value) for $M_{\rm p} = 60\,
M_\oplus$ (\citealt{dong_etal_2017}). Elsewhere within the gap,
$\Sigd/\Sigg \ll 10^{-2}$. Such subsolar metallicities appear inconsistent with most of the data plotted in Figure \ref{fig:dust}.

The apparent disagreement between theory and observation could 
arise from some systematic error in $M_{\rm p}$ or $h/r$ ($\Sigg \propto M_{\rm p}^{-1}(h/r)^4$ according to equation \ref{eqn:backward}). It could also point to an error in the simulations' assumed initial condition of solar dust abundance. 
We think the most likely explanation is that the empirical values of $\Sigd$ are overestimated because the dust gaps are underresolved. Figure \ref{fig:dust} suggests the measurement error in $\Sigd$ is at least a factor of 10-100. Errors of this magnitude are not necessarily surprising given how steep dust gradients can be in the simulations, especially for $M_{\rm p} > 10 \,M_\oplus$
(\citealt{dong_etal_2017}, third row of their figure 8). Indeed \citet[][their figure 12]{jennings_etal_2021} re-analyzed the DSHARP data and found deeper bottoms to the dust gaps in GW Lup, SR 4, AS 209-G1, HD 143006-G1, and HD 163296-G2, in many instances by orders of magnitude relative to \citet{huang_etal_2018}.

\begin{figure*} 
\includegraphics[width=\textwidth]{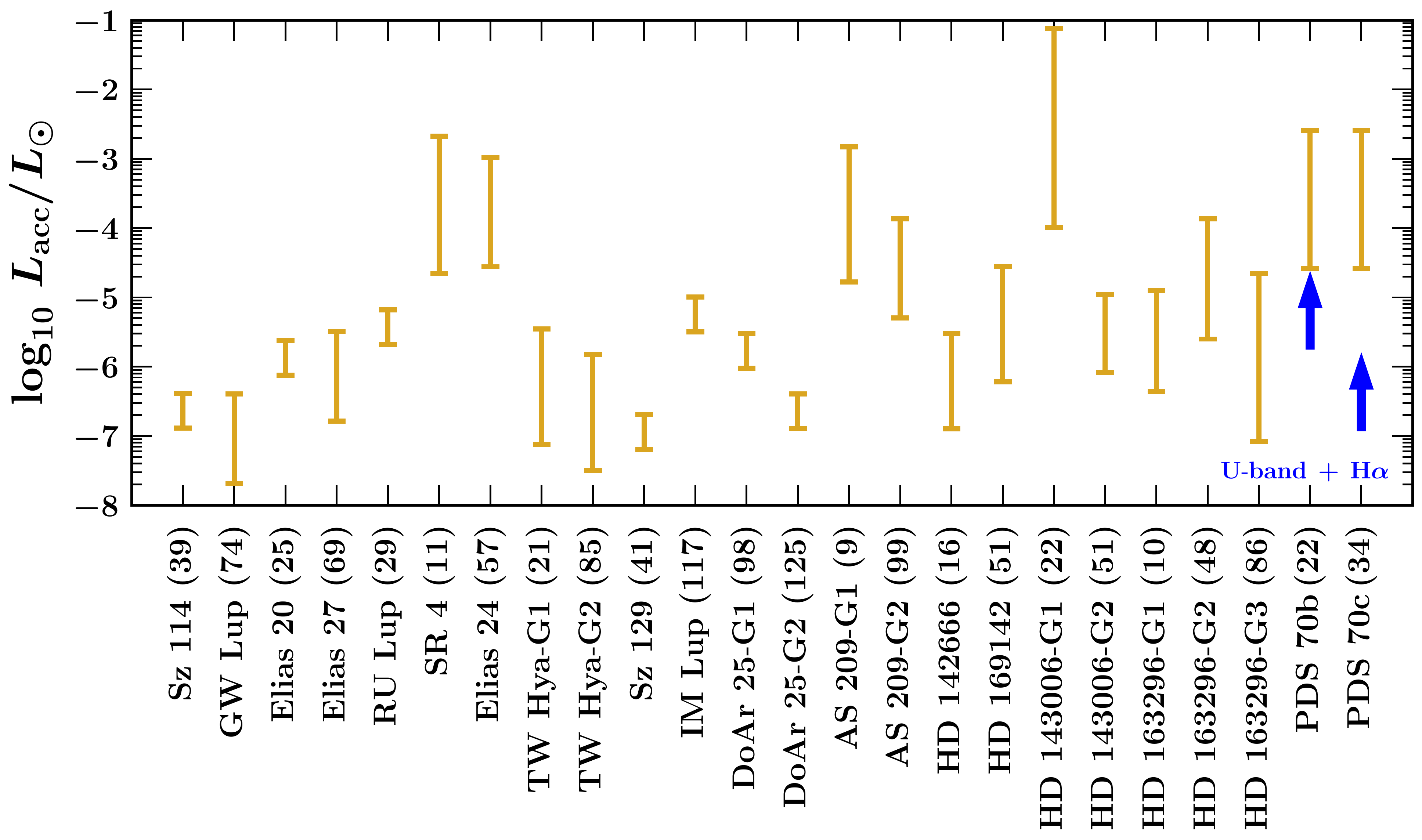}
\vspace{-2mm}
\caption{Predicted bolometric accretion luminosities $L_{\rm acc} = GM_{\rm p}\dot{M}_{\rm p}/R_{\rm p}$ assuming each planet is accreting at the average rate given by its mass and system age $\dot{M}_{\rm p} = M_{\rm p}/t_{\rm age}$ (equivalently $t_{\rm double} = t_{\rm age}$). For the non-PDS 70 planets we estimate $R_{\rm p} = R_{\rm J} (M_{\rm p}/M_{\rm J})^{1/3}$ for $M_{\rm p} < M_{\rm J}$ and $R_{\rm p} = R_{\rm J} (M_{\rm p}/M_{\rm J})^{-1/3}$ for $M_{\rm p} > M_{\rm J}$. For PDS 70b and c we use $R_{\rm p} = 2\,R_{\rm J}$ following blackbody fits by \protect \citet{wang_etal_2020} to near-infrared photometry. Error bars reflect the adopted range of $M_{\rm p}$ (see Table \ref{tab:data} or the top panel of Figure \ref{fig:bondi_prediction_gas}), including the dependence of $R_{\rm p}$ on $M_{\rm p}$. \nick{The blue arrows show estimates of $L_{\rm acc}$ for the PDS 70 planets derived from observed U-band (336 nm) and H$\alpha$ fluxes. These values are hard lower limits because the accretion fluxes may be attenuated by dust along the line of sight, or the accretion luminosity may be concentrated in wavebands other than those observed. Figures \ref{fig:sed2} and \ref{fig:sed} showing the SEDs for PDS 70b and c suggest much of the accretion power ultimately emerges in the infrared.}
}
  \label{fig:Lacc_prediction}
\end{figure*}

\begin{figure} 
\includegraphics[width=\columnwidth]{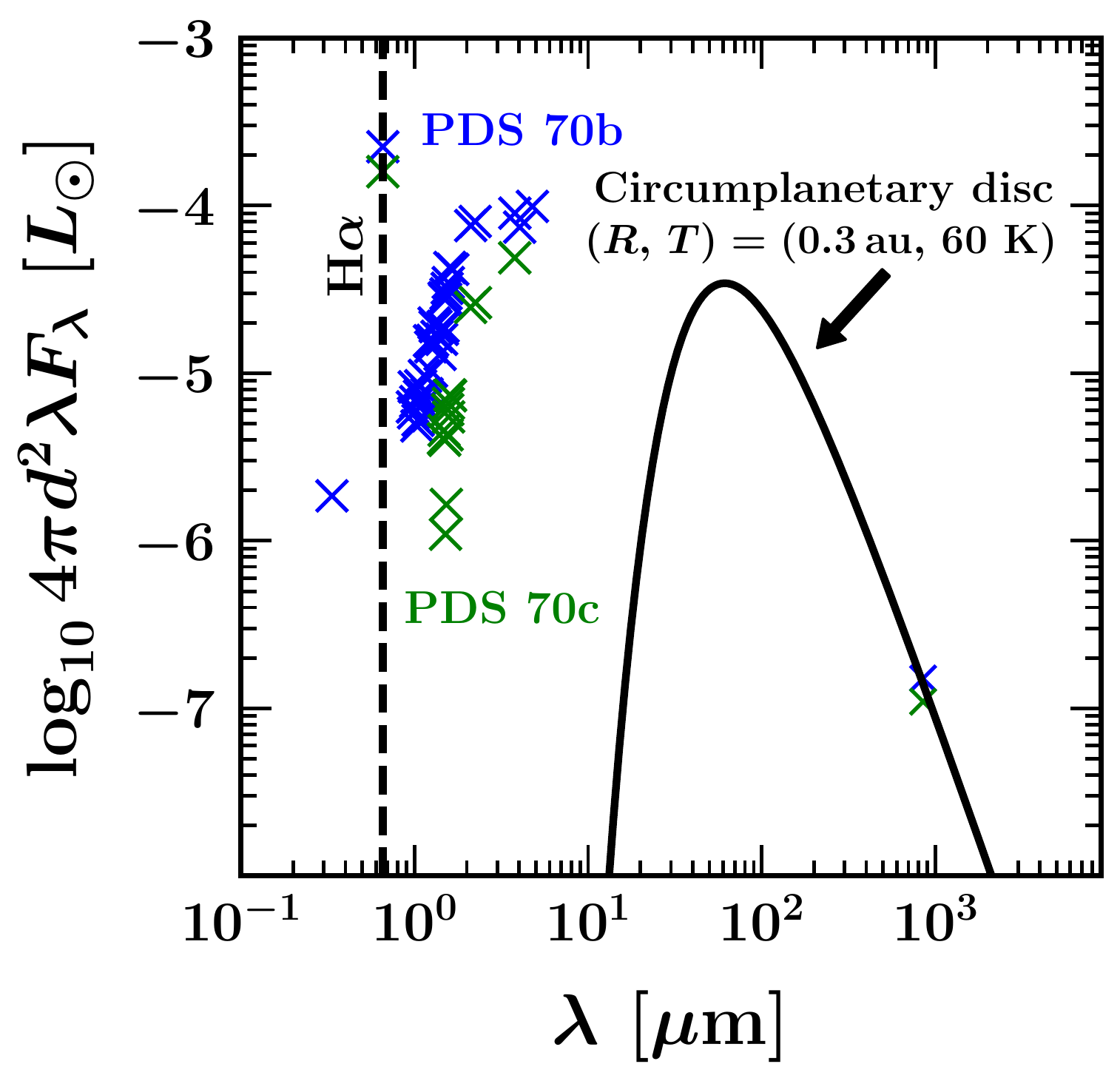}
\caption{Observed SEDs of the PDS 70 planets (crosses) and emission
  from a single-temperature blackbody modeling a circumplanetary disc
  (CPD, black curve). Wavelength is denoted by $\lambda$, flux density
  by $F_\lambda$ (units of energy per time per area per wavelength),
  and $d = 113$ pc is the source distance. The observed SEDs range from
  the U-band continuum and H$\alpha$ line \protect
  \citep{hashimoto_etal_2020, zhou_etal_2021}, to the near-infrared
  \protect \citep{muller_etal_2018, mesa_etal_2019,
    wang_etal_2020, stolker_etal_2020, wang_etal_2021}, to the sub-mm \protect \citep{isella_etal_2019, benisty_etal_2021}. The CPD temperature equals that of the local background nebula (60 K; section \ref{subsec:hr}) and is assumed to be viewed face-on with a radius $R$ chosen to match the observed mm-wave
fluxes. Mass accretion rates
calculated assuming Bondi accretion yield ultraviolet/optical accretion
luminosities of order $10^{-4}\,L_{\odot}$, much higher than implied
by the observed U-band flux (leftmost cross). Most of the accretional energy
may be released instead as Lyman-$\alpha$ photons \protect \citep{aoyama_etal_2020, aoyama_etal_2021}. 
} 
  \label{fig:sed2}
\end{figure}

\begin{figure*} 
\includegraphics[width=\textwidth]{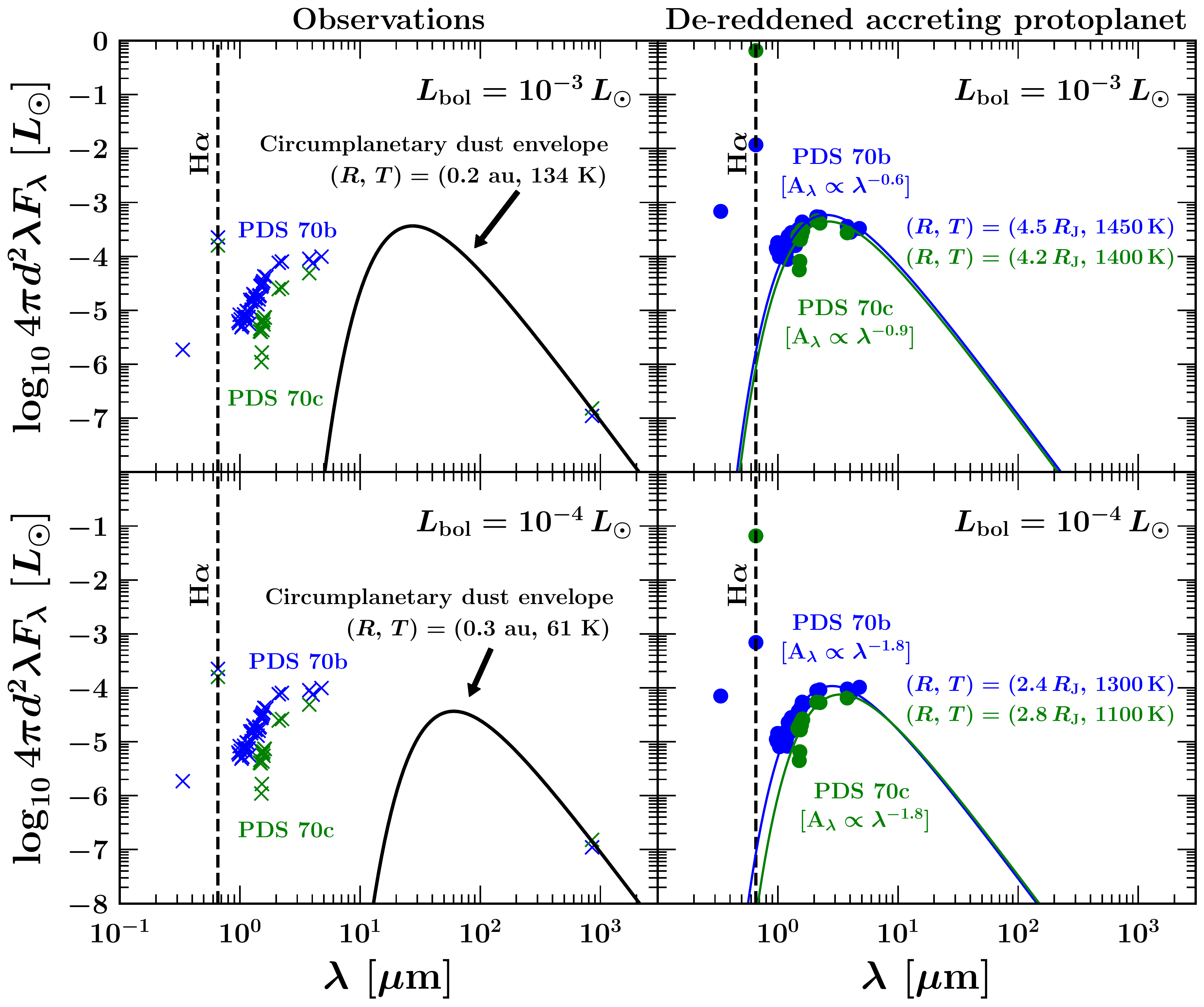}
\caption{Similar to Figure \ref{fig:sed2}, but now modeling the case where the PDS 70 planets are surrounded by dusty spherical envelopes instead of circumplanetary discs. Crosses in the left panels mark the observed SEDs, identical to those in Figure \ref{fig:sed2}. In the right panels we show these data de-reddened, adopting a power-law extinction curve chosen so that the extinction-corrected luminosity in the U-band + H$\alpha$ is comparable to that in the near-infrared, with both summing to a given bolometric luminosity $L_{\rm bol}$ ($10^{-4} L_\odot$ for the bottom panels, $10^{-3}L_\odot$ for the top). These choices are intended to describe protoplanets which derive a large fraction of their luminosity from ongoing accretion, producing short-wavelength ultraviolet/optical radiation in an accretion shock that heats the planetary layers below and the circumplanetary envelope above. The blackbody fits to the de-reddened near-infrared SEDs (right panels) suggest the accretion shock lies just above $\sim$2--5 $R_{\rm J}$. The circumplanetary envelope intercepts nearly all of the outgoing radiation and re-radiates it longward of 10--30 $\mu$m wavelength on size scales comparable to the Hill sphere, as shown by the blackbodies with luminosities equal to $L_{\rm bol}$ running through the mm-wave fluxes (left panels). The large de-reddened U-band and optical luminosities in this figure ($10^{-4}$, $10^{-3} L_\odot$) are consistent with planetary accretion at Bondi-like rates and mass doubling times $t_{\rm double}$ comparable to the system age $t_{\rm age}$.
}
  \label{fig:sed}
\end{figure*}
\subsection{Accretion luminosities, and the spectral energy distributions of the PDS 70 planets}
\label{subsec:observables}
\nick{From the average accretion rate $\dot{M}_{\rm p} = M_{\rm p}/t_{\rm age}$ we can predict the bolometric accretion luminosities of planets in our sample \citep[e.g.][]{frank_king_raine}:
\begin{equation}
    L_{\rm acc} = \frac{GM_{\rm p}\dot{M}_{\rm p}}{R_{\rm p}}.
\end{equation}
For the non-PDS 70 planets, we take radii $R_{\rm p}$ to be $R_{\rm J}(M_{\rm p}/M_{\rm J})^{-1/3}$ for $M_{\rm p} > M_{\rm J}$ and $R_{\rm J}(M_{\rm p}/M_{\rm J})^{1/3}$ for $M_{\rm p} < M_{\rm J}$. These scalings apply to planets that have finished cooling and contracting \citep[e.g.][]{ginzburg_chiang_2019b} and therefore probably underestimate 
$R_{\rm p}$. We fix $R_{\rm p} = 2\,R_{\rm J}$ for the PDS 70 planets, following fits by \cite{wang_etal_2020} to the observed near-infrared photometry.\footnote{\citet{wang_etal_2020} derived these radii by fitting blackbodies to the near-infrared spectral energy distributions of PDS 70b and c assuming no extinction. The radii may change by order-unity factors if there is circumplanetary extinction, as proposed later in this section.} 
%EC: sentence below is not supported by the figure
%    which shows instead 1e-8 to 1e-1. "Typically"
%    has no meaning when I look at the figure.
%The predicted accretion luminosities shown in Figure \ref{fig:Lacc_prediction} are typically $\sim$$10^{-6}-10^{-4}\,L_{\odot}$. 
}

\nick{Figure \ref{fig:Lacc_prediction} also marks in blue the observed luminosities of PDS 70b and c at the short wavelengths thought to characterize the hot accretion shock. For PDS 70b, \cite{zhou_etal_2021} estimated $L_{\rm acc} = 1.8 \times 10^{-6}\,L_{\odot}$ from the measured U-band (336 nm) and H$\alpha$ luminosities. This value includes a bolometric correction for the hydrogen continua, calculated using a slab model \citep{valenti_etal_1993} having parameters similar to those estimated for the accretion shock \citep{aoyama_ikoma_2019}. PDS 70c has so far only been detected in H$\alpha$ emission \citep{hashimoto_etal_2020}, so we conservatively set $L_{\rm acc} = L_{\rm H\alpha} = 1.2 \times 10^{-7}\,L_{\odot}$. Figure \ref{fig:Lacc_prediction} shows that these luminosities (in blue) sit a factor of $\sim$100 below our estimates based on the time-averaged $\dot{M}_{\rm p}$ (in yellow).}

\nick{There are at least two ways to reconcile our theorized high accretion luminosities for PDS 70b and c with the low short-wavelength luminosities observed to date. One possibility is that the bulk of the accretion luminosity emerges in as yet unobserved wavebands. In the accretion shock model of \cite{aoyama_etal_2021}, most of the power comes out in the hydrogen Lyman-$\alpha$ emission line. Their model for PDS 70b predicts a total accretion luminosity $L_{\rm acc} \approx 5\times 10^{-5} - 5\times 10^{-4} L_\odot$ (see their figure 1). Alternatively, the observed U-band and H$\alpha$ fluxes may be highly attenuated by dust and therefore underestimate $L_{\rm acc}$ when not corrected for extinction \citep{hashimoto_etal_2020}. 
These two explanations are not mutually exclusive; reality could be some combination of high dust extinction and high Lyman-$\alpha$ luminosity.
In either scenario the true $L_{\rm acc}$ could be $\gtrsim 10^{-4}\,L_\odot$, compatible with our Bondi accretion rate estimates.}

\nick{We illustrate in Figures \ref{fig:sed2} and \ref{fig:sed} how these two possibilities impact our interpretation of the spectral energy distributions (SEDs) of PDS 70b and c. Figure \ref{fig:sed2} describes the no-extinction case. Here the short-wavelength luminosity generated from the accretion shock is $L_{\rm acc} \sim 10^{-4}L_\odot$, released mostly in some unobserved waveband, possibly H Ly$\alpha$ (\citealt{aoyama_etal_2021}). That the observed luminosity in the near-infrared (see tables 3 and 4 of \citealt{wang_etal_2020}) is comparable to our predicted $L_{\rm acc}$ may not be a coincidence. Most of the near-infrared power may derive from the half of the radiation generated in the shock that is emitted toward and re-processed by cooler $\mathcal{O}(10^3\,\rm K)$ layers below the shock (see \citealt{aoyama_etal_2020} for a model that details how the accretion power is re-processed).}

In this extinction-free scenario the observed fluxes from the ultraviolet to the near-infrared are not significantly extincted by dust. \nick{At the same time we know the circumplanetary environment should contain dust to explain the mm-wave continuum emission imaged by 
\cite{isella_etal_2019} and \citet[][]{benisty_etal_2021}.} For this mm-bright dust to not obscure our line of sight to the planet, it should be distributed in a circumplanetary disc (CPD), viewed face on or approximately so. Figure \ref{fig:sed2} shows blackbody emission from a single-temperature CPD, heated primarily by radiation from the star to an assumed $T = 60$ K, with a radius fitted to the mm-wave observations. The fitted CPD radius is 0.3 au, roughly 20\% of the Hill sphere radius $R_{\rm Hill} \sim  1.5\,\,\mathrm{au} \,(M_{\rm p}/M_{\rm J})^{1/3}(r/22\,\rm au)$.

Figure \ref{fig:sed} shows a high-extinction scenario.  \citet{wang_etal_2020} ruled out interstellar extinction because the host star appears unextincted, and circumstellar extinction seems negligible because the planets reside within a transitional disc cavity. Thus we are left with circumplanetary extinction, from dust brought to the planet by the nebular accretion flow, pervading the planet's Hill sphere. To obscure our line of sight to the accretion shock, this dust should be distributed more-or-less spherically around the planet, in a geometry similar to the thick torii found in 3D hydro-simulations by \citet[][see their figure 2]{fung_etal_2019}. The left panels of Figure \ref{fig:sed} demonstrate how Hill-sphere-sized dusty  spheres/torii can reproduce the measured sub-mm fluxes. 

\nick{The right panels of Figure \ref{fig:sed} show what the extinction-corrected intrinsic SED of the accreting planet might look like. We adopt power laws for how the extinction $A_{\lambda}$ varies with wavelength $\lambda$, with the slope and normalization chosen such that the extinction-corrected luminosity in the U-band + H$\alpha$ is comparable to the extinction-corrected luminosity in the near-infrared, with both summing to an assumed bolometric luminosity $L_{\rm bol}$ ($10^{-4} L_\odot$ for the bottom panels, $10^{-3} L_\odot$ for the top; see the figure annotations and caption for details). Our V-band extinction for PDS 70b is $A_{\rm V} \approx 2.5$ for $L_{\rm bol} = 10^{-4} L_\odot$, and $A_{\rm V} \approx 4$ for $L_{\rm bol} = 10^{-3}\,L_{\odot}$. The de-reddened near-infrared SEDs still conform approximately to blackbodies, about as well as they do without any extinction correction applied (cf. \citealt{wang_etal_2020}). Our extinction-corrected effective temperatures are higher than non-corrected values, $\sim$1100--1450 K vs. 1000--1200 K. Our corrected planet radii are also larger, 2.4--4.5 $R_{\rm J}$ vs. 2--2.7 $R_{\rm J}$. In this interpretation the extinction-corrected near-infrared radius corresponds to the depth at which radiation emitted by the accretion shock boundary layer thermalizes, at the base of a dusty accretion flow. The analogy would be with actively accreting Class 0 protostars (e.g. \citealt{dunham_etal_2014}) or ``first'' or ``second'' protostellar cores (\citealt{larson_1969}; \citealt{bate_etal_2014}) at the centres of infalling dusty envelopes.}

\section{Summary and Outlook}
\label{sec:discussion}

Planets are suspected to open 
the gaps and cavities imaged in 
protoplanetary discs at
millimeter and infrared wavelengths.
From the literature we have compiled the modeled masses $M_{\rm p}$ and orbital radii $r$ of the
putative gap-opening planets, together with local disc properties 
including the dust and gas surface densities $\Sigd$ and $\Sigg$, and
the gas disc thickness $h$. Our sample contains 22
hypothesized planets and
2 confirmed orbital companions (PDS 70b and c). From these data we have evaluated:

\begin{enumerate}
\item Present-day planetary gas accretion rates $\dot{M}_{\rm p}$ 
for the subset of 9 planets (including PDS 70b and c) for which the ambient gas density can be usefully constrained from molecular line observations including optically thin C$^{18}$O. We assumed, consistent with some theories of planet formation, that accretion from the surrounding disc onto the planet is Bondi-like. 
For 8 of the 9 planets, accretion rates are such
that mass doubling times $t_{\rm double} \equiv M_{\rm p}/\dot{M}_{\rm p}$ 
are comparable to stellar ages of $t_{\rm age} \sim 1-10$ Myr --- this assumes the non-PDS 70 planets have masses $M_{\rm p} \sim 10-30 \,\Mearth$, the PDS 70 planets have $M_{\rm p} \sim 1-10\,\Mj$, and the gas-phase CO:H$_2$ abundance ratios are close to those found in  thermo-chemical models.
 Since circumstellar gas discs have typical lifetimes of several Myr, our finding that $t_{\rm double}/t_{\rm age} \sim 1$ for these 8 planets suggests we are observing them during their last doublings. The one planet that does not fit this pattern is the one supposedly inside HD 163296-G2 ($r = 48$ au), for which we find $t_{\rm double}/t_{\rm age} \lesssim 0.1$ if $M_{\rm p} \sim 1-10\,M_{\rm J}$ as estimated by \cite{zhang_etal_2018}. 
We suggest the true planet mass in this gap is smaller, and could actually be zero if the gap is created by a planet located elsewhere, following \citet{dong_etal_2018}.

\item Gas surface densities $\Sigg$ at the bottoms of 24 gaps where planets are supposed to reside. For 9 systems we compiled literature estimates of $\Sigg$ based on observed C$^{18}$O intensity maps. For the remaining 15 systems we inferred $\Sigg$ assuming the gaps host planets undergoing Bondi accretion at the average rate $M_{\rm p}/t_{\rm age}$. The predicted gas surface densities range from $10^{-4}\,\gcm$ to $1\,\gcm$.

\item Dust surface densities $\Sigd$ and dust-to-gas ratios $\Sigd/\Sigg$ inside gaps. In many cases $\Sigd/\Sigg$ based  on observations are supersolar, by contrast to the typically subsolar ratios found at the bottoms of dust-filtered gaps in planet-disc simulations (\citealt{dong_etal_2017}). We suggest $\Sigd$ may be overestimated by the current ALMA observations as they may not be resolving steep dust gradients inside gaps (see also \citealt{jennings_etal_2021}). Under-resolution is less of a problem for $\Sigg$ insofar as simulations predict gas gradients to be shallower than dust gradients.
\item \nick{Accretion luminosities $L_{\rm acc}$ of protoplanets. For the PDS 70 planets we predict $L_{\rm acc} \sim 10^{-4}\,L_{\odot}$, a factor of 10-100 larger than observed in the U-band and H$\alpha$. Most of the short-wavelength accretion power might 
%be hidden in 
be emitted in wavebands as yet unobserved, e.g. H Lyman-$\alpha$  as predicted by recent accretion shock models \citep{aoyama_etal_2021}. Alternatively, a dusty, quasi-spherical, circumplanetary  envelope might
be absorbing 
%attenuate nearly all 
much 
of the outgoing ultraviolet/optical radiation. We sketched different spectral
energy distributions (SEDs)
based on these scenarios. A large fraction of the power is expected to emerge at wavelengths ranging from the mid to far-infrared.}

\end{enumerate}

\nick{The hypothesis that the disc 
substructures imaged by ALMA are 
caused by embedded planets is now supported on several grounds.
A planet can reproduce the observed positions of multiple gaps in a given system (e.g. in AS 209; \citealt{dong_etal_2017, dong_etal_2018}; \citealt{zhang_etal_2018}). A planet can also reproduce observed  non-Keplerian velocity signatures in the disc rotation curve (e.g. in HD 163296; \citealt{teague_etal_2018, teague_etal_2021, pinte_etal_2020}). To this  evidence favoring planets we can now add 
consistency with planet accretion theory. The data are pointing to Neptune-mass planets accreting at Bondi-like rates inside %annular
gaps filled with low-viscosity ($\alpha \lesssim 10^{-4}$) gas.}

From here there are any number of avenues for future investigation. To list just a few:

  \begin{enumerate}
   \item Advancing beyond our Bondi-inspired formula (\ref{eqn:Mdot_bondi}) for $\dot{M}_{\rm p}$. Bondi accretion neglects
many effects, among them stellar tides that become important when the planet's Bondi radius exceeds its Hill radius (e.g. \citealt{rosenthal_etal_2020}), anisotropic 3D flow fields (e.g. \citealt{cimerman_etal_2017}), and 
thermodynamics (e.g. \citealt{ginzburg_chiang_2019a} who distinguished between hydrodynamic and thermodynamic runaway). The sensitivity of simulated accretion rates to sink-cell (e.g. \citealt{dangelo_etal_2003}) and other planetary boundary conditions should be studied.

\item Accounting for how the gap density evolves with time because of disc photoevaporation, repulsive planetary torques, etc., while the planet accretes. For example, in the gap clearing theories of \citet{ginzburg_chiang_2019a}, the planet mass doubling timescale $t_{\rm double}$ 
is longer than the instantaneously measured $M_{\rm p}/\dot{M}_{\rm p}$ by a factor that ranges up to $\sim$15, increasing as the disc viscosity decreases and the gap depletes more rapidly with time. \nick{This correction factor based on disk evolution should be incorporated into comparisons between $t_{\rm double}$ and $t_{\rm age}$.}

\item More molecular line data to test our predictions for $\Sigg$, and higher spatial resolution to map steep $\Sigd$ profiles. New data are now available for the GM Aur disc, including spatially resolved dust continuum and CO observations (\citealt{huang_etal_2020, huang_etal_2021}; \citealt{schwarz_etal_2021}), anchored by a total disc gas mass estimate from HD emission \citep{mcclure_etal_2016}. \nick{Spatially resolved disc surveys have so far targeted brighter discs \citep{andrews_etal_2018, oberg_etal_2021} and are consequently biased toward higher $\Sigg$.
Assuming planets will always
be found with $t_{\rm double}/t_{\rm age} \gtrsim 1$, higher $\Sigg$ corresponds to lower planet masses $M_{\rm p}$ (equation \ref{eqn:backward}). 
By this logic, targeting less luminous, less massive discs while keeping all other factors fixed  may uncover higher mass planets. In any case expanding survey samples are necessary for understanding planet occurrence rates and demographic trends \citep{cieza_etal_2019}.}

\item \nick{Planetary accretion shock models \citep{aoyama_etal_2020, aoyama_etal_2021} need to be tested observationally across the ultraviolet-optical spectrum, with current arguments based on a single emission line (H$\alpha$) replaced by joint constraints from multiple lines and the continuum. The accretion shock theory also needs to connect more smoothly to models of the cooler, infrared-emitting layers of the planet (e.g. BT-Settl; \citealt{baraffe_etal_2015}); at the moment how radiation from shocked layers heats cooler layers is not accounted for (but see section A1 of \citealt{aoyama_etal_2020} for a first step). Detection of molecular absorption lines in the infrared \citep{cugno_etal_2021, wang_etal_2021} may constrain the degree of accretional heating, and how much dust is brought in. }

\end{enumerate}

\section*{Acknowledgements}
We thank Yuhiko Aoyama, Steve Beckwith, Jenny Calahan, Yayaati Chachan, Ruobing (Robin) Dong, Jeffrey Fung, Sivan Ginzburg, Jane Huang, Masahiro Ikoma, Gabriel-Dominique Marleau, Jason Wang, Ke (Coco) Zhang, Shangjia Zhang, Yifan Zhou, and Zhaohuan Zhu, for discussions and feedback on our study. We also thank Takayuki Muto for a helpful referee report. This work was supported by an NSF Graduate Research Fellowship (DGE 2146752) and used the \textsc{matplotlib} \citep{hunter_etal_2007} and \textsc{scipy} \citep{scipy_2020} packages.

\section*{Data availability}
The data compiled in this work are available upon request to the authors.

%%%%%%%%%%%%%%%%%%%%%%%%%%%%%%%%%%%%%%%%%%%%%%%%%%



\bibliographystyle{mnras}
\bibliography{planets_nick} 

%%%%%%%%%%%%%%%%%%%%%%%%%%%%%%%%%%%%%%%%%%%%%%%%%%

%%%%%%%%%%%%%%%%% APPENDICES %%%%%%%%%%%%%%%%%%%%%

%%%%%%%%%%%%%%%%%%%%%%%%%%%%%%%%%%%%%%%%%%%%%%%%%%

% Don't change these lines
\bsp	% typesetting comment
\label{lastpage}
\end{document}